\documentclass[journal,technote]{IEEEtran}

\usepackage[T1]{fontenc}
\usepackage[utf8]{inputenc}
\usepackage{textcomp}
\usepackage{bm}

\usepackage{amsmath,amssymb,amsfonts}

\usepackage{graphicx}
\usepackage[caption=false,font=footnotesize]{subfig}
\usepackage{array}
\usepackage{stfloats}
\usepackage{float}
\usepackage{xcolor}
\usepackage{booktabs}

\usepackage{algorithm}
\usepackage{algorithmic}

\usepackage{cite}
\usepackage{url}

\usepackage{verbatim}
\usepackage{enumitem}
\usepackage{soul}

\usepackage{cleveref}

\crefname{figure}{Fig.}{Figs.}
\AtBeginDocument{
    \setlength{\abovedisplayskip}{4pt plus 1pt minus 1pt}
    \setlength{\belowdisplayskip}{4pt plus 1pt minus 1pt}
    \setlength{\abovedisplayshortskip}{2pt plus 1pt minus 1pt}
    \setlength{\belowdisplayshortskip}{2pt plus 1pt minus 1pt}
    \setlength{\jot}{1pt}
}

\begin{document}

\title{Joint LSFC and Nakagami-\(m\) Estimation From Downlink SINR in GZF-Precoded CF-mMIMO}

\author{Longfei~Liu,
        Yichen~Guo,
        Lili~Xu, 
        Yihang~Cheng,
        Tao~Peng,~\IEEEmembership{Member,~IEEE,}
        and~Wenbo~Wang,~\IEEEmembership{Senior~Member,~IEEE}

\thanks{This work has been submitted to the IEEE for possible publication.
Copyright may be transferred without notice, after which this version may
no longer be accessible.}
\thanks{This work was supported in part by the National Science and Technology Major Project - Mobile Information Networks (No. 2026ZD1307800) and in part by the BUPT-China Unicom Joint Innovation Center (No. 2025-STHZ-BJYDDX-007). \textit{(Corresponding author: Tao Peng)}}
\thanks{ Longfei Liu, Tao Peng and Wenbo Wang are with the Key Laboratory of Universal Wireless Communications, Ministry of Education, Beijing University of Posts and Telecommunications, Beijing 100876, China (e-mail: 6longfei@bupt.edu.cn;  pengtao@bupt.edu.cn; wbwang@bupt.edu.cn).}
\thanks{ Yichen Guo is with the School of Physics and Electronic Information Engineering, Minjiang University, Fuzhou 350108, China (e-mail: guoyichen@bupt.edu.cn).}
\thanks{ Lili Xu and Yihang Cheng are with the Research Institute of China United Network Communications Co., Ltd, Beijing 100176, China (e-mail: xull82@chinaunicom.cn; chengyh60@chinaunicom.cn).}
}

\markboth{IEEE TRANSACTIONS ON VEHICULAR TECHNOLOGY,~Vol.~XX, No.~X, XXX~2026}%
{Liu \MakeLowercase{\textit{et al.}}: Joint LSFC and Nakagami-\(m\) Estimation From Downlink SINR in GZF-Precoded CF-mMIMO}


\maketitle

\begin{abstract}
Accurate characterization of the instantaneous signal-to-interference-plus-noise ratio (SINR) is important for reliability-sensitive cell-free massive multiple-input multiple-output (CF-mMIMO) systems. This paper proposes a joint estimation method for link-specific large-scale fading coefficients (LSFCs) and a common Nakagami-\(m\) shape parameter using power and downlink SINR measurements. A tractable SINR distribution is first derived under global zero-forcing (GZF) precoding, Nakagami-\(m\) fading, and link-level fractional power control (FPC). Based on the derived probability density function (PDF), a model-driven neural network (NN) is developed to solve the resulting maximum-likelihood estimation (MLE) problem. Simulations show that the reconstructed SINR distributions obtained with the estimated parameters closely match the empirical distributions. With \(2\times10^{5}\) samples, the proposed method achieves an LSFC root-mean-square error (RMSE) of approximately \(1.0\)~dB and an \(m\) RMSE of approximately \(0.085\).
\end{abstract}


\section{Introduction}
\IEEEPARstart{T}{o} enhance macro-diversity and service uniformity, cell-free massive multiple-input multiple-output (CF-mMIMO) enables coherent cooperation among geographically distributed access points (APs) without relying on conventional cell boundaries~\cite{Ngo2017CellFree,Nayebi2017Precoding}. These advantages make CF-mMIMO attractive for reliability-sensitive IMT-2030 services, including immersive communication and hyper-reliable and low-latency communication (HRLLC)~\cite{ITU2023M2160}. For such services, neither average spectral efficiency nor average signal-to-interference-plus-noise ratio (SINR) can adequately characterize rare link failures, which are governed by the lower tail of the instantaneous downlink SINR distribution~\cite{Bennis2018URLLC,Lancho2023URLLC}. Therefore, characterizing this distribution is essential for assessing link reliability.

From the propagation perspective, the instantaneous SINR distribution is shaped by both large-scale and small-scale fading parameters. The link-specific large-scale fading coefficients (LSFCs) determine the average strengths of the AP--user equipment (UE) links, whereas the small-scale fading law governs their instantaneous fluctuations. In CF-mMIMO, highly nonuniform AP--UE gains can cause only a few APs to dominate, limiting channel hardening and preserving the influence of small-scale fading on the instantaneous SINR~\cite{Chen2018Hardening,Interdonato2019Training}. We therefore adopt the Nakagami-\(m\) model for non-line-of-sight (NLoS) links to capture varying fading severities~\cite{VanTorre2011IndoorMIMO}.

Existing CF-mMIMO studies mainly formulate SINR characterization as a forward problem. Outage probabilities and SINR distributions have been derived under Rayleigh fading by treating the LSFCs and power control coefficients as known inputs~\cite{Chong2024SINRDistribution,Huang2025EEEURLLC}. These forward analyses do not address the inverse recovery of propagation parameters from instantaneous SINR observations. Data-driven LSFC mining supports interference identification and SINR prediction but does not estimate fading severity, whereas a prior SINR-based estimator jointly estimates the Nakagami-\(m\) parameter and link-specific uplink interference but is not formulated for CF-mMIMO~\cite{Chen2023LSFCMining,Liu2026MLE}. To the best of our knowledge, joint estimation of link-specific LSFCs and fading severity from instantaneous downlink SINR and recorded link-level powers under global zero-forcing (GZF) and varying fractional power control (FPC) remains unaddressed.

In this work, we develop a framework for jointly estimating these propagation parameters based on the downlink SINR distribution. The main contributions are as follows:

\begin{itemize}

\item We derive tractable approximations for the probability density function (PDF) and cumulative distribution function (CDF) of the instantaneous downlink SINR under GZF precoding, Nakagami-\(m\) fading, and link-level FPC.

\item Based on the derived SINR PDF, we formulate the joint estimation of the link-specific LSFCs and the Nakagami-\(m\) parameter from SINR and power measurements as a maximum-likelihood estimation (MLE) problem and develop a model-driven neural network (NN) to solve it.

\item Monte Carlo simulations show close agreement between the reconstructed SINR CDFs obtained with the estimated parameters and the corresponding empirical CDFs.

\end{itemize}

Sections II--IV present the system model, SINR analysis, and proposed estimator, respectively. Section V presents simulation results, and Section VI concludes the paper.

\section{System Model}

We consider a time-division duplex (TDD) CF-mMIMO system where \(N\) geographically distributed \(M\)-antenna APs jointly serve \(K\) single-antenna UEs over shared time-frequency resources. A central processing unit (CPU) performs channel processing and downlink precoding. Each coherence block comprises uplink training followed by downlink data transmission, with channels constant throughout.

Let
\(\bm h_{ik}=\sqrt{\beta_{ik}}\bm g_{ik}\in\mathbb C^{M\times1}\)
denote the noise-normalized channel from AP \(i\) to UE \(k\), where \(l_{ik}\) and
\(\sigma^2\) are the physical large-scale channel gain and receiver
noise power, respectively, and \(\beta_{ik}=l_{ik}/\sigma^2\) is the
noise-normalized LSFC. We model scattering-dominated NLoS or partially obstructed links using Nakagami-\(m\) fading~\cite{VanTorre2011IndoorMIMO} and focus on
\(m\in[0.5,1.5]\). For antenna \(r\), \(|g_{ik,r}|^2\sim\mathcal G(m,1/m)\), with a phase independent of the amplitude and uniformly distributed over \([0,2\pi)\), where \(\mathcal G(\alpha,\theta)\) denotes the Gamma distribution with shape \(\alpha\) and scale \(\theta\). The coefficients \(g_{ik,r}\) are mutually independent across \(i\), \(k\), \(r\), and are independently realized across coherence blocks, whereas \(\beta_{ik}\) and \(m\) remain fixed during data collection. Thus, \(|h_{ik,r}|^2\sim\mathcal G(m,\beta_{ik}/m)\). For parsimony, we retain link-specific LSFCs but assume a common \(m\) across all AP--UE links, as also assumed for multiple Nakagami branches in~\cite{Lin2021NakagamiCombiner}.

\subsection{Uplink Training}

The UEs transmit mutually orthogonal pilots of length
\(n_{\mathrm{ul}}\geq K\) with uplink power \(p^{\mathrm{ul}}\).
After correlating the received pilot signal with UE \(k\)'s pilot,
AP \(i\) obtains
\begin{equation}
    \bm y_{ik}
    =
    \sqrt{q_{\mathrm p}}\bm h_{ik}+\bm n_{ik},
    \qquad
    q_{\mathrm p}\triangleq n_{\mathrm{ul}}p^{\mathrm{ul}},
\end{equation}
where \(\bm n_{ik}\sim\mathcal{CN}(\bm0,\bm I_M)\) is the normalized noise. The linear minimum mean-square error (LMMSE) estimate is
\(\hat{\bm h}_{ik}=c_{ik}\bm y_{ik}\), and the estimation error is
\(\tilde{\bm h}_{ik}=\bm h_{ik}-\hat{\bm h}_{ik}\), where
\begin{equation}
c_{ik}=\frac{\sqrt{q_{\mathrm p}}\beta_{ik}}
{q_{\mathrm p}\beta_{ik}+1},\quad
\nu_{ik}=\frac{q_{\mathrm p}\beta_{ik}^{2}}
{q_{\mathrm p}\beta_{ik}+1},\quad
\zeta_{ik}=\beta_{ik}-\nu_{ik}.
\end{equation}

Accordingly,
\(\mathbb E[\hat{\bm h}_{ik}\hat{\bm h}_{ik}^{H}]
=\nu_{ik}\bm I_M\) and
\(\mathbb E[\tilde{\bm h}_{ik}\tilde{\bm h}_{ik}^{H}]
=\zeta_{ik}\bm I_M\). By the LMMSE orthogonality principle,
\(\mathbb E[\hat{\bm h}_{ik}\tilde{\bm h}_{ik}^{H}]=\bm0\).
For \(m\neq1\), the estimate and error are generally non-Gaussian and dependent despite this orthogonality.

\subsection{Power Control and GZF Precoding}

Let \(\mathcal U_i\subseteq\{1,\ldots,K\}\) denote the set of UEs served by AP \(i\), and let \(\mathcal U=\bigcup_{i=1}^{N}\mathcal U_i\) denote the active UE set. Let \(\eta_{ij}\geq0\) denote the power control coefficient selected by the CPU, where \(\eta_{ij}=0\) for \(j\notin\mathcal U_i\) and \(\sum_{j\in\mathcal U_i}\eta_{ij}\leq1\). With a downlink power budget \(p^{\mathrm{dl}}\) at each AP, define \(p_{ij}\triangleq{p}^{\mathrm{dl}}\eta_{ij}\) and collect these values in \(\bm P=[p_{ij}]\in\mathbb R_{+}^{N\times K}\). The applied \(\bm P\) is recorded by the CPU and treated as known in the subsequent estimation. The specific FPC rule used to generate the simulation configurations is given in Section V.

The estimated channel of UE \(j\) is
\(\hat{\bm h}_j
=[\hat{\bm h}_{1j}^{T},\ldots,\hat{\bm h}_{Nj}^{T}]^{T}\),
and
\(\hat{\bm H}=[\hat{\bm h}_j]_{j\in\mathcal U}\).
Assuming \(NM\geq|\mathcal U|\) and that \(\hat{\bm H}\) has full
column rank, the unnormalized GZF precoding matrix and the normalized
precoding vector for UE \(j\) are
\begin{equation}
\bar{\bm V}
=
\hat{\bm H}
(\hat{\bm H}^{H}\hat{\bm H})^{-1},
\qquad
\bm v_j
=
\frac{\bar{\bm v}_j}{\|\bar{\bm v}_j\|},
\end{equation}
respectively, where \(\bar{\bm v}_j\) is the column of
\(\bar{\bm V}\) corresponding to UE \(j\). The global precoding vector
is partitioned as
\(\bm v_j=[\bm v_{1j}^{T},\ldots,\bm v_{Nj}^{T}]^{T}\), where
\(\bm v_{ij}\in\mathbb C^{M\times1}\) denotes the precoding segment
applied at AP \(i\).

Define the power loading matrix for UE \(j\) as
\begin{equation}
\bm Q_j
\triangleq
\operatorname{diag}
\big(
\sqrt{p_{1j}}\bm I_M,\ldots,
\sqrt{p_{Nj}}\bm I_M
\big).
\end{equation}

Thus, \(\bm v_j\) is the normalized GZF beam direction, while \(\bm Q_j\bm v_j\) is the effective precoding vector after power loading.

The signal transmitted by AP \(i\) is \(\bm s_i
=
\sum_{j\in\mathcal U_i}
\sqrt{p_{ij}}\bm v_{ij}x_j\), where the data symbols are independent with zero mean and unit power. Its average transmit power satisfies
\begin{equation}
\mathbb E[\|\bm s_i\|^2]
=
\sum_{j\in\mathcal U_i}
p_{ij}\|\bm v_{ij}\|^2
\leq
\sum_{j\in\mathcal U_i}p_{ij}
\leq
p^{\mathrm{dl}}.
\end{equation}

Using \(\bm h_k=\hat{\bm h}_k+\tilde{\bm h}_k\), the received signal
at UE \(k\in\mathcal U\) is expressed as
\begin{equation}
y_k^{\mathrm{dl}}
=
\hat{\bm h}_k^{H}\bm Q_k\bm v_kx_k
+\sum_{\substack{j\in\mathcal U\\j\neq k}}
\hat{\bm h}_k^{H}\bm Q_j\bm v_jx_j
+
\sum_{j\in\mathcal U}
\tilde{\bm h}_k^{H}\bm Q_j\bm v_jx_j
+n_k,
\end{equation}
where \(n_k\sim\mathcal{CN}(0,1)\). The four terms represent the desired signal, leakage caused by unequal power loading, interference due to estimation errors, and receiver noise, respectively. The GZF directions satisfy \(\hat{\bm h}_k^{H}\bm v_j=0\) for \(j\neq k\). Unequal power loading across APs generally yields \(\bm Q_j\not\propto\bm I\), making \(\hat{\bm h}_k^{H}\bm Q_j\bm v_j\) potentially nonzero.

\section{SINR Distribution Analysis}

Following the SINR decomposition in~\cite{Huang2025EEEURLLC},
the instantaneous downlink SINR of UE \(k\in\mathcal U\) is given by
\begin{equation}
\gamma_k
=
\frac{
\left|\hat{\bm h}_k^{H}\bm Q_k\bm v_k\right|^2
}{
\displaystyle
\sum_{\substack{j\in\mathcal U\\j\neq k}}
\left|\hat{\bm h}_k^{H}\bm Q_j\bm v_j\right|^2
+
\sum_{j\in\mathcal U}
\left|\tilde{\bm h}_k^{H}\bm Q_j\bm v_j\right|^2
+1
}
=
\frac{DS_k}{INT_k}.
\end{equation}

\subsection{Desired Signal Approximation}

The exact distribution of \(DS_k\) is difficult to obtain because the normalized GZF beam direction depends on the inverse Gram matrix of all active UEs and its AP blocks are unequally loaded by \(\bm Q_k\). We therefore
derive a tractable approximation using Gamma moment matching, a Schur
complement, a diagonal Gram approximation, and the Delta method.

\emph{1) Channel Moments and Schur Representations:}
Under the assumed Nakagami-\(m\) fading model and uniform phase distribution,
\(\mathbb E[h_{ik,r}^{2}]=0\) and
\(\mathbb E[|h_{ik,r}|^{4}]
=\beta_{ik}^{2}(1+1/m)\).
Substituting
\(\hat h_{ik,r}
=c_{ik}(\sqrt{q_{\mathrm p}}h_{ik,r}+n_{ik,r})\)
and matching the first two moments of
\(|\hat h_{ik,r}|^2\) gives
\begin{equation}
\begin{aligned}
|\hat h_{ik,r}|^2
&\sim
\mathcal G\left(
\hat\kappa_{ik},
\frac{\nu_{ik}}{\hat\kappa_{ik}}
\right),
\hat\kappa_{ik}
=
\frac{(q_{\mathrm p}\beta_{ik}+1)^2}
{(q_{\mathrm p}\beta_{ik}+1)^2+
q_{\mathrm p}^{2}\beta_{ik}^{2}(1/m-1)}.
\end{aligned}
\label{eq:estimated_power_gamma}
\end{equation}

For the target UE \(k\), define the estimated channel energy moments for AP \(i\) as
\begin{equation}
\mu_{ik}
\triangleq
\mathbb E\!\left[\|\hat{\bm h}_{ik}\|^2\right]
=M\nu_{ik},
\
\omega_{ik}
\triangleq
\operatorname{Var}\!\left(\|\hat{\bm h}_{ik}\|^2\right)
=
\frac{M\nu_{ik}^{2}}{\hat\kappa_{ik}}.
\label{eq:channel_energy_moments}
\end{equation}

Let \(\mathcal U_{-k}\triangleq\mathcal U\setminus\{k\}\),
let \(\hat{\bm H}_{-k}\) collect the corresponding estimated channels,
and denote the corresponding Gram matrix by
\(\bm G_{-k}\triangleq
\hat{\bm H}_{-k}^{H}\hat{\bm H}_{-k}\).
The Schur complement gives
\begin{equation}
\begin{aligned}
\bm\Pi_{-k}^{\perp}
&\triangleq
\bm I_{NM}
-\hat{\bm H}_{-k}\bm G_{-k}^{-1}\hat{\bm H}_{-k}^{H},\\
S_k
&=\hat{\bm h}_k^{H}\bm\Pi_{-k}^{\perp}\hat{\bm h}_k,
\quad
T_k=\hat{\bm h}_k^{H}\bm Q_k\bm\Pi_{-k}^{\perp}\hat{\bm h}_k,
\quad
DS_k=\frac{|T_k|^2}{S_k}.
\end{aligned}
\label{eq:schur_representation}
\end{equation}

Since \(\|\bar{\bm v}_k\|^2=S_k^{-1}\), \(S_k\) accounts for
GZF normalization, while \(T_k\) represents the projected channel
gain after power loading.

\emph{2) Moment Approximation:}
Motivated by favorable propagation, we approximate the inverse Gram matrix by the reciprocals of its mean diagonal entries~\cite{Ngo2014Favorable}. This approximation is intended for the strongly underloaded regime \(NM\gg|\mathcal U|\):

\begin{equation}
\bm G_{-k}^{-1}
\approx
\operatorname{diag}
\left(
\frac{1}{\Lambda_j}
\right)_{j\in\mathcal U_{-k}},
\
\Lambda_j
\triangleq
\mathbb E\!\left[\|\hat{\bm h}_{j}\|^2\right]
=
M\sum_{i=1}^{N}\nu_{ij}.
\label{eq:gram_inverse_approx}
\end{equation}

Substituting \eqref{eq:gram_inverse_approx} into the Schur representation and replacing the projection terms involving other UEs by their conditional means given \(\hat{\bm h}_k\) gives
\begin{equation}
a_{ik}\triangleq\sqrt{p_{ik}},
\
s_{ik}\triangleq
1-\sum_{j\in\mathcal U_{-k}}\frac{\nu_{ij}}{\Lambda_j},
\
t_{ik}\triangleq a_{ik}s_{ik}.
\label{eq:sti_definition}
\end{equation}

This gives
\begin{equation}
S_k
\approx
\sum_{i=1}^{N}s_{ik}\|\hat{\bm h}_{ik}\|^2,
\
T_k
\approx
\sum_{i=1}^{N}t_{ik}\|\hat{\bm h}_{ik}\|^2.
\label{eq:leading_st_representation}
\end{equation}

Using the channel energy moments in \eqref{eq:channel_energy_moments} yields
\begin{equation}
\begin{aligned}
\mu_{S,k}
&=\sum_{i=1}^{N}s_{ik}\mu_{ik},
&
\mu_{T,k}
&=\sum_{i=1}^{N}t_{ik}\mu_{ik},\\
v_{S,k}^{(0)}
&=\sum_{i=1}^{N}s_{ik}^{2}\omega_{ik},
&
v_{T_{\mathrm R},k}^{(0)}
&=\sum_{i=1}^{N}t_{ik}^{2}\omega_{ik},\\
c_{ST,k}^{(0)}
&=\sum_{i=1}^{N}s_{ik}t_{ik}\omega_{ik}.
\end{aligned}
\label{eq:basic_st_moments}
\end{equation}

The diagonal Gram approximation neglects residual fluctuations in the
projection terms involving other UEs. To account for these fluctuations,
define, for
\(j\in\mathcal U_{-k}\) and \(\ell\in\{0,1,2\}\),
\(
e_{\ell,j}
\triangleq
\sum_{i=1}^{N}a_{ik}^{\ell}\mu_{ik}\nu_{ij},
\
r_{\ell,j}
\triangleq
\sum_{i=1}^{N}a_{ik}^{\ell}\omega_{ik}\nu_{ij}^{2}.
\)
The dependence of \(e_{\ell,j}\) and \(r_{\ell,j}\) on the target
UE \(k\) is omitted for notational brevity.

Writing \(T_k=T_{\mathrm R,k}+\mathrm jT_{\mathrm I,k}\), we have
\(\mathbb E[S_k]\approx\mu_{S,k}\),
\(\mathbb E[T_{\mathrm R,k}]\approx\mu_{T,k}\), and
\(\mathbb E[T_{\mathrm I,k}]\approx0\).
Let \(v_{S,k}\), \(v_{T_{\mathrm R},k}\), and
\(v_{T_{\mathrm I},k}\) denote the corresponding variances, and let
\(c_{ST,k}\triangleq\operatorname{Cov}(S_k,T_{\mathrm R,k})\).
The corrected second-order moments are
\begin{equation}
\begin{aligned}
v_{S,k}
&=
v_{S,k}^{(0)}
+
\sum_{j\in\mathcal U_{-k}}
\frac{
\bigl(e_{0,j}\bigr)^2+r_{0,j}
}
{\Lambda_j^{2}},\\
v_{T_{\mathrm R},k}
&=
v_{T_{\mathrm R},k}^{(0)}
+
\frac{1}{2}
\sum_{j\in\mathcal U_{-k}}
\frac{
e_{0,j}e_{2,j}
+\bigl(e_{1,j}\bigr)^2
+2r_{2,j}
}
{\Lambda_j^{2}},\\
v_{T_{\mathrm I},k}
&=
\left[
\frac{1}{2}
\sum_{j\in\mathcal U_{-k}}
\frac{
e_{0,j}e_{2,j}
-\bigl(e_{1,j}\bigr)^2
}
{\Lambda_j^{2}}
\right]^+,\\
c_{ST,k}
&=
c_{ST,k}^{(0)}
+
\sum_{j\in\mathcal U_{-k}}
\frac{
e_{0,j}e_{1,j}
+r_{1,j}
}
{\Lambda_j^{2}}.
\end{aligned}
\label{eq:corrected_st_moments}
\end{equation}

Here, \([x]^+\triangleq\max(x,0)\) removes small negative values
caused by numerical roundoff.

\emph{3) Gamma Approximation:}
Since \(DS_k=|T_k|^2/S_k\) is a nonlinear function of the jointly
fluctuating \(S_k\), \(T_{\mathrm R,k}\), and \(T_{\mathrm I,k}\), its
first two moments are approximated using the Delta method
\cite{Oehlert1992Delta}:
\begin{equation}
\begin{aligned}
\mu_{DS,k}
&\approx
\frac{\mu_{T,k}^{2}}{\mu_{S,k}}
+
\frac{v_{T_{\mathrm R},k}+v_{T_{\mathrm I},k}}{\mu_{S,k}}
-
\frac{2\mu_{T,k}c_{ST,k}}{\mu_{S,k}^{2}}
+
\frac{\mu_{T,k}^{2}v_{S,k}}{\mu_{S,k}^{3}},\\
\sigma_{DS,k}^{2}
&\approx
\frac{\mu_{T,k}^{4}v_{S,k}}{\mu_{S,k}^{4}}
+
\frac{4\mu_{T,k}^{2}v_{T_{\mathrm R},k}}{\mu_{S,k}^{2}}
-
\frac{4\mu_{T,k}^{3}c_{ST,k}}{\mu_{S,k}^{3}}
+
\frac{2v_{T_{\mathrm I},k}^{2}}{\mu_{S,k}^{2}}.
\end{aligned}
\label{eq:x_delta_moments}
\end{equation}

We match the resulting nonnegative variable to a Gamma distribution
using its first two moments
\cite{Chong2024SINRDistribution,Huang2025EEEURLLC}:
\begin{equation}
DS_k
\sim
\mathcal G(\alpha_{k,1},\theta_{k,1}),
\
\alpha_{k,1}
=
\frac{\mu_{DS,k}^{2}}{\sigma_{DS,k}^{2}},
\
\theta_{k,1}
=
\frac{\sigma_{DS,k}^{2}}{\mu_{DS,k}}.
\label{eq:x_gamma_approx}
\end{equation}

\subsection{Interference Approximation}

Define the leakage caused by unequal power loading and the interference due to estimation errors as
\[
I_{k\leftarrow j}
=
|\hat{\bm h}_k^H\bm Q_j\bm v_j|^2,
\ j\in\mathcal U_{-k},
\
E_{k\leftarrow j}
=
|\tilde{\bm h}_k^H\bm Q_j\bm v_j|^2,
\ j\in\mathcal U.
\]

They are approximated as
\(I_{k\leftarrow j}\sim\mathcal G(1,q_{k\leftarrow j}^{I})\) and
\(E_{k\leftarrow j}\sim\mathcal G(1,q_{k\leftarrow j}^{E})\), where
\begin{equation}
\begin{aligned}
q_{k\leftarrow j}^{I}
&=
\frac{NM}{NM-1}
\frac{
\sum_{i=1}^{N}
\bigl(\sqrt{p_{ij}}-\bar a_{kj}\bigr)^2
\nu_{ik}\nu_{ij}
}{
\sum_{i=1}^{N}\nu_{ij}
},\\
q_{k\leftarrow j}^{E}
&=
\frac{
\sum_{i=1}^{N}p_{ij}\zeta_{ik}\nu_{ij}
}{
\sum_{i=1}^{N}\nu_{ij}
},
\qquad
\bar a_{kj}
=
\frac{
\sum_{i=1}^{N}\sqrt{p_{ij}}\,\nu_{ik}
}{
\sum_{i=1}^{N}\nu_{ik}
}.
\end{aligned}
\label{eq:interference_scales}
\end{equation}

The scales in \eqref{eq:interference_scales} use second moments and the mean energy of each AP block in the normalized GZF beam direction. For \(q_{k\leftarrow j}^{E}\), the residual dependence between \(\tilde{\bm h}_k\) and \(\bm v_j\) is neglected and the unconditional error covariance \(\mathbb E[\tilde{\bm h}_{ik}\tilde{\bm h}_{ik}^{H}]=\zeta_{ik}\bm I_M\) is used; hence, it is not an exact conditional scale.

The leakage scale \(q_{k\leftarrow j}^{I}\) vanishes under uniform loading across the AP blocks of \(\bm v_j\), but is generally nonzero under unequal loading.

For tractability, covariances among different interference components are neglected. Therefore, the first two moments of \(INT_k\) are
\begin{equation}
\begin{aligned}
    \mu_{INT,k}
    &=
    1+
    \sum_{\substack{j\in\mathcal U\\j\neq k}}
    q_{k\leftarrow j}^{I}
    +
    \sum_{j\in\mathcal U}
    q_{k\leftarrow j}^{E},\\
    \sigma_{INT,k}^{2}
    &=
    \sum_{\substack{j\in\mathcal U\\j\neq k}}
    (q_{k\leftarrow j}^{I})^2
    +
    \sum_{j\in\mathcal U}
    (q_{k\leftarrow j}^{E})^2.
\end{aligned}
\end{equation}

Moment matching then gives
\(INT_k\sim\mathcal G(\alpha_{k,2},\theta_{k,2})\), where
\(
    \alpha_{k,2}
    =
    \mu_{INT,k}^2/\sigma_{INT,k}^2,
    \
    \theta_{k,2}
    =
    \sigma_{INT,k}^2/{\mu_{INT,k}}.
\)

\subsection{SINR PDF and CDF}
Neglecting the weak dependence between \(DS_k\) and \(INT_k\),
we approximate them as independent Gamma random variables. The PDF
of \(\gamma_k=DS_k/INT_k\) is then
\begin{equation}
    f_{\gamma_k}(\gamma)
    =
    \frac{
    \Gamma(\alpha_{k,1}+\alpha_{k,2})
    \gamma^{\alpha_{k,1}-1}
    \theta_{k,1}^{\alpha_{k,2}}
    \theta_{k,2}^{\alpha_{k,1}}
    }
    {
    \Gamma(\alpha_{k,1})
    \Gamma(\alpha_{k,2})
    \left(
    \theta_{k,1}+\gamma\theta_{k,2}
    \right)^{\alpha_{k,1}+\alpha_{k,2}}
    }.
\label{eq:sinr_pdf}
\end{equation}

Integrating the PDF yields the corresponding CDF
\begin{equation}
    F_{\gamma_k}(\gamma)
    =
    I_{
    \frac{\gamma\theta_{k,2}}
    {\theta_{k,1}+\gamma\theta_{k,2}}
    }
    \left(
    \alpha_{k,1},
    \alpha_{k,2}
    \right),
\label{eq:sinr_cdf}
\end{equation}
where \(I_z(a,b)\) denotes the regularized incomplete beta function.

\section{Problem Formulation and Proposed Algorithm}

\subsection{Dataset Acquisition}

Following the system model in Section II, for each UE \(k\), the CPU
records \(E\) measurement entries over independently realized coherence
blocks~\cite{Chen2023LSFCMining}. Entry \(e\) contains the matrix of recorded link-level powers
\(\bm P^{(e)}=[p_{ij}^{(e)}]\in\mathbb R_{+}^{N\times K}\)
and the observed instantaneous SINR
\(\gamma_{k,e}^{\mathrm{obs}}\), where \(p_{ij}^{(e)}\) is the power
assigned by AP \(i\) to UE \(j\), with \(p_{ij}^{(e)}=0\) indicating
an inactive link. The resulting dataset is
\(
\mathcal D_k
=
\left\{
\left(\bm P^{(e)},\gamma_{k,e}^{\mathrm{obs}}\right)
\right\}_{e=1}^{E}.
\)

\subsection{Problem Formulation}

We jointly estimate the link-specific noise-normalized LSFCs and a
common Nakagami-\(m\) shape parameter. Let
\(\bm\beta=[\beta_{ij}]\in\mathbb R_{+}^{N\times K}\). The unknown
parameter set is
\begin{equation}
\boldsymbol{\Theta}
=
\{\bm\beta,m\}.
\label{eq:parameter_set}
\end{equation}

For each entry \(e\), \(\bm P^{(e)}\) is known, while
\(\gamma_{k,e}^{\mathrm{obs}}\) is converted from dB to linear scale.
Given a candidate \(\boldsymbol{\Theta}\), each likelihood contribution is evaluated using the GZF-based conditional SINR PDF approximation \(f_{\gamma_k}(\cdot)\) in
\eqref{eq:sinr_pdf}. The negative log-likelihood (NLL) over all UE
datasets is
\begin{equation}
\mathcal L(\boldsymbol{\Theta})
=
-
\sum_{k=1}^{K}
\sum_{e=1}^{E}
\log
f_{\gamma_k}
\left(
\gamma_{k,e}^{\mathrm{obs}}
\mid
\bm P^{(e)};\boldsymbol{\Theta}
\right).
\label{eq:nll_objective}
\end{equation}

The joint estimation problem is formulated as
\begin{equation}
\hat{\boldsymbol{\Theta}}
\triangleq
\{\hat{\bm\beta},\hat m\}
=
\arg\min_{\substack{\beta_{ij}>0,\ \forall i,j\\ m\ge 1/2}}
\mathcal L(\boldsymbol{\Theta}).
\label{eq:joint_estimation}
\end{equation}

\subsection{Proposed Algorithm}

To approximately solve \eqref{eq:joint_estimation}, the proposed
estimator adopts a model-driven multiple-input single-output
NN-assisted MLE (NN-MLE) architecture without conventional hidden
layers for feature extraction. In each mini-batch, the recorded link-level powers and the observed
SINRs constitute the multiple data inputs, while the system
configuration is known and fixed. Each
link-level input \(p_{ij}^{(e)}\) is paired with the corresponding
trainable LSFC \(\beta_{ij}\), whereas \(m\) is shared across all
links. Fixed analytical layers compute the LMMSE statistics and the Gamma
parameters of \(DS_k\) and \(INT_k\), and then evaluate the
conditional SINR PDF at the observed samples. The resulting sample
NLL terms are aggregated at a single scalar output node to form the
mini-batch loss. Backpropagation through the analytical model
updates \(\bm\beta\) and \(m\). Unlike a predictor trained offline, NN-MLE is reoptimized for each acquired dataset \(\{\mathcal D_k\}_{k=1}^{K}\).

For numerical stability, the LSFCs are optimized in the dB domain and
converted to linear scale during likelihood evaluation. We employ alternating updates across UEs. With \(m\) fixed, each column
of \(\bm\beta\) is successively updated using the mini-batch NLL
associated with \(\mathcal D_k\), while the remaining columns are
fixed. Since \(f_{\gamma_k}\) depends on the full \(\bm\beta\), each
update uses the current values of all other columns, and repeated
sweeps propagate their coupling. After each sweep over the LSFC
columns, \(\bm\beta\) is fixed and \(m\) is updated by cycling through
all UE datasets. The procedure terminates when the relative change in the NLL falls
below a prescribed threshold or the maximum number of outer
iterations is reached.

\subsection{Computational Complexity}

Let \(E\) denote the number of entries per UE. Evaluating the desired
signal moments and interference scales for one target UE costs
\(O(ENK)\) per epoch. Using \(I_{\beta}\) epochs for each LSFC
column and \(I_m\) epochs over all UE datasets for \(m\) in each of
\(I_{\mathrm{out}}\) outer iterations requires
\(O(I_{\mathrm{out}}(I_{\beta}+I_m)ENK^2)\).

\section{Simulation Results}

\subsection{Simulation Scenario} 

We consider a CF-mMIMO network over a square area, where \(N\) APs and \(K\) UEs are independently and uniformly deployed, with each AP equipped with \(M\) antennas. The noise-normalized LSFC is generated in dB as \(
10\log_{10}(\beta_{ik})
=
-30.5-36.7\log_{10}(d_{ik})+F_{ik}
-P_{\sigma^2,\mathrm{dBm}},
\)
where \(d_{ik}\) is the three-dimensional AP--UE distance, \(F_{ik}\) denotes shadow fading, and \(P_{\sigma^2,\mathrm{dBm}}\) is the noise power. For simulation data generation, these LSFCs are used to construct the power control coefficients according to the following FPC rule:
\begin{equation}
\eta_{ij}
=
\begin{cases}
\displaystyle
\frac{\beta_{ij}^{\chi}}
{\sum_{\ell\in\mathcal U_i}\beta_{i\ell}^{\chi}},
& j\in\mathcal U_i,\\[2mm]
0, & j\notin\mathcal U_i,
\end{cases},
\end{equation}
where \(p_{ij}=p^{\mathrm{dl}}\eta_{ij}\). Within each estimation
dataset, the FPC exponent \(\chi\) takes 25 values in \([-3,4]\),
yielding diverse \(\bm P^{(e)}\), whereas SINR CDF validation uses a
fixed \(\chi\) for each setting; \(\chi=0\) gives equal power among the
active links at each AP. Small-scale fading follows a Nakagami-\(m\)
distribution with \(m\in[0.5,1.5]\) for scattering-dominated NLoS
links~\cite{VanTorre2011IndoorMIMO}.

\begin{table}[b]
\centering
\caption{Simulation Parameters}
\label{tab:simulation_parameters}
\footnotesize
\setlength{\tabcolsep}{4.5pt}
\renewcommand{\arraystretch}{0.90}
\begin{tabular}{@{}c|c@{}}
\hline
Parameter & Value \\
\hline
Network area & $100\,\mathrm{m}\times100\,\mathrm{m}$ \\
Numbers of APs, UEs, and antennas per AP \((N,K,M)\)
& \((15,6,10)\) \\
Height difference between AP and UE & 10 m \\
Bandwidth $B$ & 20 MHz \\
Pilot sequence length $n_{\mathrm{ul}}$ & 20 \\
Uplink transmit power $p^{\mathrm{ul}}$ & 100 mW \\
Downlink power budget per AP \(p^{\mathrm{dl}}\) & 200 mW \\
Noise power \(P_{\sigma^2,\mathrm{dBm}}\) & \(-94\) dBm \\
Standard deviation of shadow fading & 4 dB \\
True Nakagami-\(m\) shape parameter \(m\) & \([0.5,1.5]\) \\
\hline
\end{tabular}
\end{table}

\subsection{Estimation and Distribution Accuracy}

\begin{figure}[t]
\centering
\subfloat[]{%
    \includegraphics[width=0.48\columnwidth]{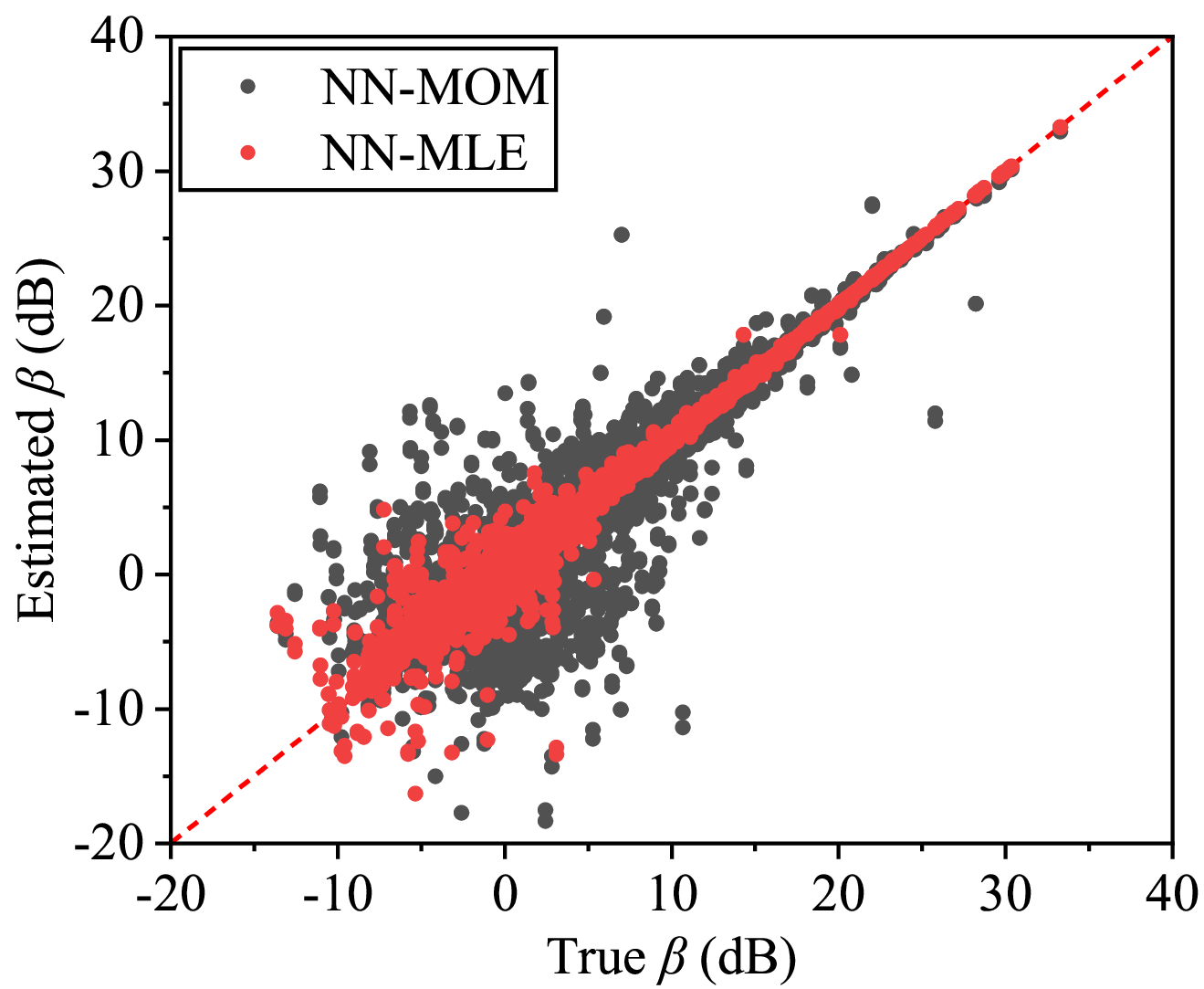}%
}\hfill
\subfloat[]{%
    \includegraphics[width=0.495\columnwidth]{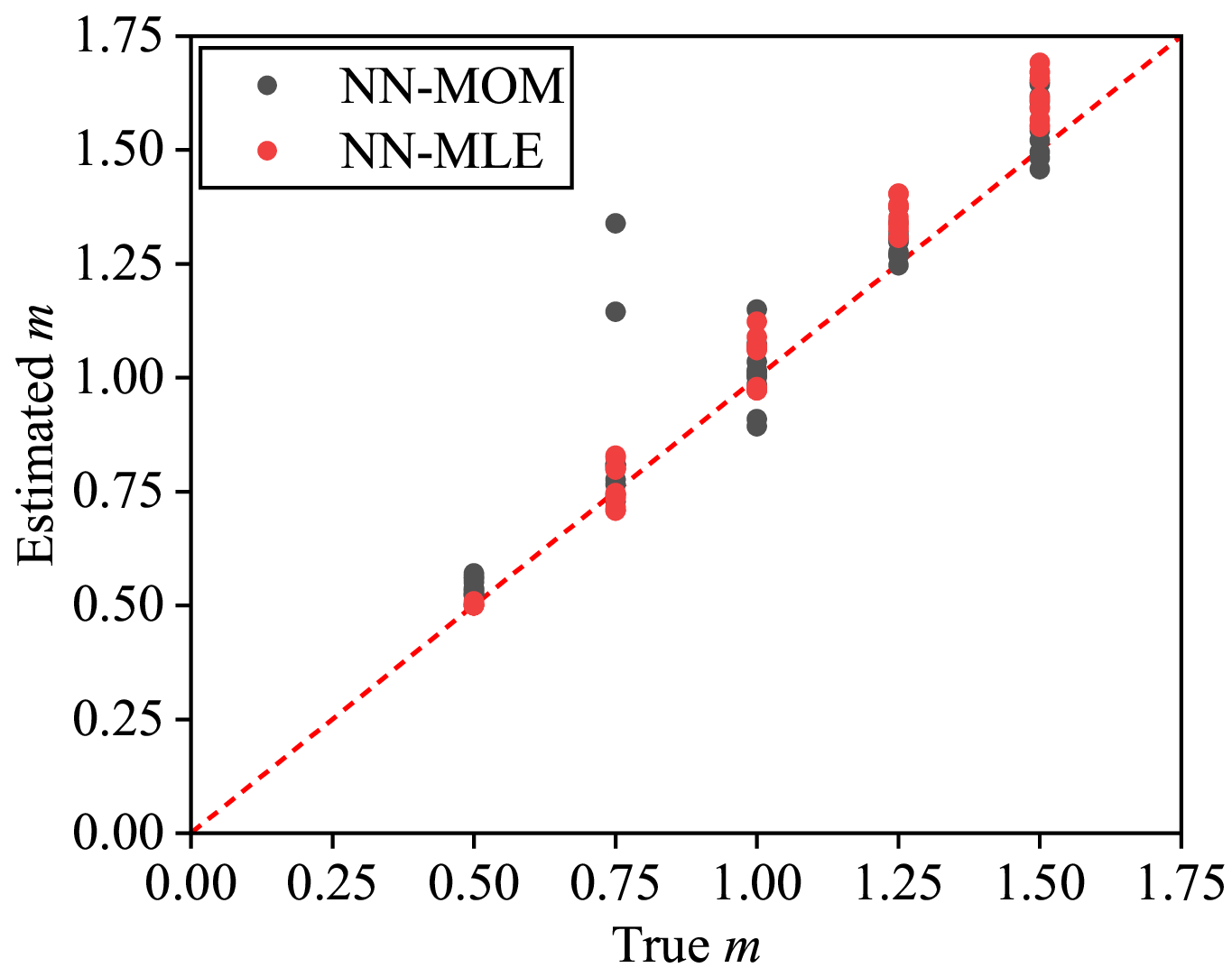}%
}
\caption{Parameter estimation performance.
(a) LSFC estimation results.
(b) Nakagami-\(m\) shape parameter estimation results.}
\label{fig:estimation}
\end{figure}

\emph{1) Parameter Estimation Performance:}
Fig.~\ref{fig:estimation} compares the proposed NN-MLE with an NN-assisted method-of-moments (NN-MOM) baseline, which matches the sample and model means and variances of grouped SINR in dB instead of using the NLL. Five true values
\(m\in\{0.5,0.75,1,1.25,1.5\}\) are considered, with ten independent
datasets per value and \(2\times10^{5}\) SINR samples per UE.
As shown in Fig.~\ref{fig:estimation}(a), NN-MLE achieves an LSFC root-mean-square error (RMSE)
of approximately \(1\) dB, compared with \(3\) dB for NN-MOM, whose
errors are particularly large for weak links. Fig.~\ref{fig:estimation}(b)
shows corresponding \(m\) RMSEs of \(0.085\) and \(0.125\), respectively.
Moreover, as the dataset size increases from \(10^{3}\) to
\(2\times10^{5}\), the LSFC and \(m\) RMSEs of NN-MLE decrease overall
from \(3.09\) to \(1.01\) dB and from \(0.34\) to \(0.085\),
respectively. These results demonstrate the advantage of likelihood
optimization over moment matching.

With \(2\times10^{5}\) SINR entries per UE, a single NN-MLE run requires on the order of \(10^{2}\) s on an NVIDIA GeForce RTX 3060, supporting its computational feasibility.

\begin{figure}[t]
\centering
\subfloat[]{%
    \includegraphics[width=0.485\columnwidth]{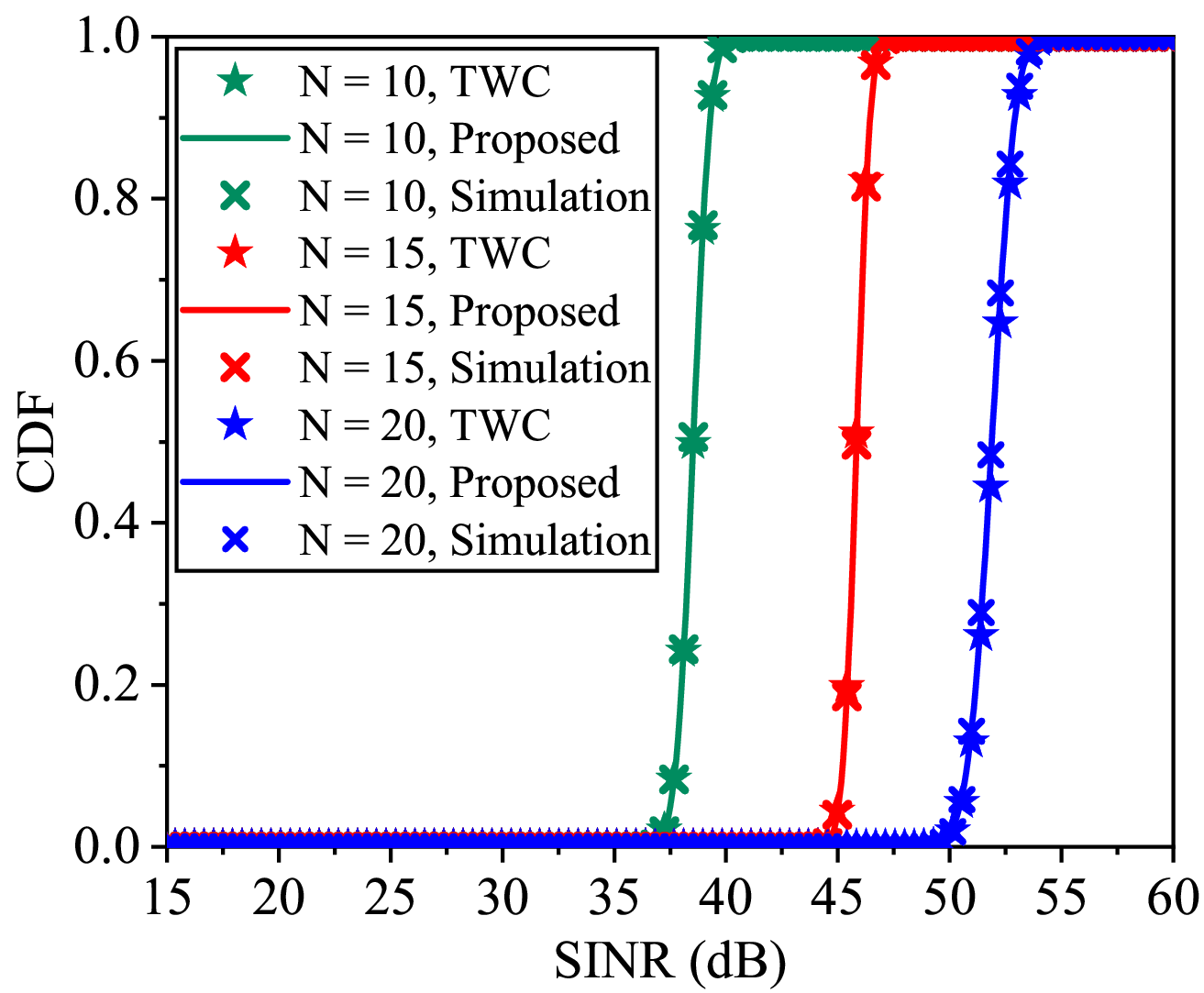}%
}\hfill
\subfloat[]{%
    \includegraphics[width=0.485\columnwidth]{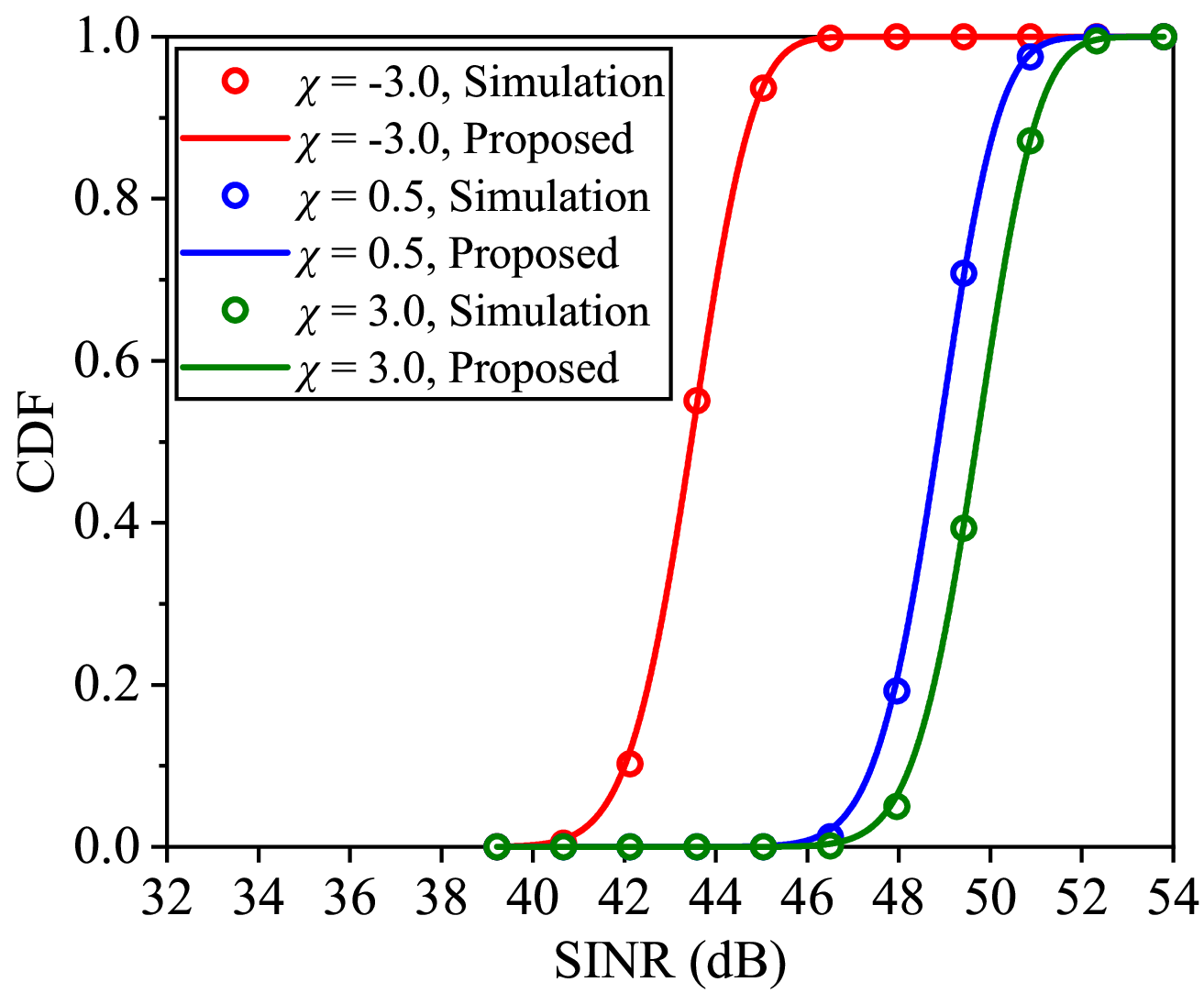}%
}
\caption{Validation of the reconstructed GZF SINR CDF.
(a) Equal-power Rayleigh case.
(b) \(m=0.5\) under different FPC exponents \(\chi\).}
\label{fig:cdf}
\end{figure}

\emph{2) SINR Distribution Accuracy:}
Fig.~\ref{fig:cdf} compares the reconstructed SINR CDF, obtained by
substituting \(\hat{\bm\beta}\) and \(\hat m\) into
\eqref{eq:sinr_cdf}, with the empirical CDF from
\(3\times10^{4}\) Monte Carlo realizations. In
Fig.~\ref{fig:cdf}(a), the equal-power Rayleigh case is considered
for different numbers of APs. Both the proposed CDF and the TWC
expression~\cite{Huang2025EEEURLLC} closely match the Monte Carlo
results, confirming consistency of the characterization with the
existing Rayleigh result. Fig.~\ref{fig:cdf}(b) considers \(m=0.5\)
for different FPC exponents \(\chi\in\{-3,0.5,3\}\). Across all UEs and
these FPC settings, the Pearson correlation coefficient satisfies
\(\left|\rho_{\mathrm P}(DS_k,INT_k)\right|<0.01\),
empirically supporting the weak dependence approximation underlying
\eqref{eq:sinr_cdf}. Despite the resulting shifts in the SINR
distribution, the reconstructed CDF remains closely aligned with the
Monte Carlo results, with Kolmogorov--Smirnov (KS) distances of \(0.0164\)--\(0.0172\) and
errors in the 1st-percentile SINR below \(0.29\) dB. These results demonstrate that the estimated parameters enable
accurate reconstruction of the SINR distribution beyond the Rayleigh
model and under diverse FPC settings.

\emph{3) Effect of Joint \(m\) Estimation:}
Although channel hardening makes the SINR distributions for different \(m\) visually similar, Table~\ref{tab:m_ablation} shows lower average LSFC RMSE under joint estimation than with fixed \(m=1\). For \(m\neq1\), this indicates that the likelihood retains information about fading severity; at \(m=1\), allowing \(m\) to vary may reshape the approximate NLL landscape and help the alternating optimizer reach a solution with lower LSFC error.

\begin{table}[t]
\caption{Effect of joint Nakagami-\(m\) estimation on LSFC recovery:
overall \(\beta\) RMSE in dB (mean \(\pm\) standard deviation).}
\label{tab:m_ablation}
\centering
\scriptsize
\setlength{\tabcolsep}{2.6pt}
\begin{tabular}{@{}lccccc@{}}
\toprule
& \multicolumn{5}{c}{True Nakagami-\(m\) shape parameter} \\
\cmidrule(lr){2-6}
Method & 0.50 & 0.75 & 1.00 & 1.25 & 1.50 \\
\midrule
Joint NN-MLE
& \textbf{1.07$\pm$0.35}
& \textbf{1.17$\pm$0.47}
& \textbf{0.90$\pm$0.29}
& \textbf{0.55$\pm$0.18}
& \textbf{0.88$\pm$0.63} \\
Fixed \(m=1\) NN-MLE
& 1.11$\pm$0.40
& 1.24$\pm$0.45
& 1.10$\pm$0.34
& 0.69$\pm$0.21
& 0.99$\pm$0.58 \\
\bottomrule
\end{tabular}
\end{table}

\section{Conclusion}

This paper derived tractable PDF and CDF approximations for the instantaneous downlink SINR of CF-mMIMO systems under GZF precoding, Nakagami-\(m\) fading, and link-level FPC. Based on the PDF approximation, a model-driven NN-MLE framework was developed to jointly estimate the link-specific noise-normalized LSFCs and a common Nakagami-\(m\) shape parameter from recorded link-level powers and instantaneous SINR observations. Numerical results showed that NN-MLE outperformed NN-MOM in parameter recovery, while the reconstructed CDFs closely matched their empirical counterparts under both Rayleigh and non-Rayleigh fading and across different FPC settings. Ablation results confirmed lower LSFC RMSE for joint \(m\) estimation than for fixed \(m=1\).

\bibliographystyle{IEEEtran}
\bibliography{references}

@article{Ngo2017CellFree,
  author  = {Hien Quoc Ngo and Alexei Ashikhmin and Hong Yang and
             Erik G. Larsson and Thomas L. Marzetta},
  title   = {Cell-Free Massive {MIMO} Versus Small Cells},
  journal = {IEEE Trans. Wireless Commun.},
  year    = {2017},
  volume  = {16},
  number  = {3},
  pages   = {1834--1850},
  month   = mar,
  doi     = {10.1109/TWC.2017.2655515}
}

@article{Nayebi2017Precoding,
  author  = {Elina Nayebi and Alexei Ashikhmin and Thomas L. Marzetta and
             Hong Yang and Bhaskar D. Rao},
  title   = {Precoding and Power Optimization in Cell-Free Massive
             {MIMO} Systems},
  journal = {IEEE Trans. Wireless Commun.},
  year    = {2017},
  volume  = {16},
  number  = {7},
  pages   = {4445--4459},
  month   = jul,
  doi     = {10.1109/TWC.2017.2698449}
}

@techreport{ITU2023M2160,
  author      = {{ITU-R}},
  title       = {Framework and Overall Objectives of the Future Development
                 of {IMT} for 2030 and Beyond},
  institution = {International Telecommunication Union},
  type        = {Recommendation},
  number      = {ITU-R M.2160-0},
  year        = {2023},
  month       = nov,
  url         = {https://www.itu.int/rec/R-REC-M.2160-0-202311-I/en}
}

@article{Bennis2018URLLC,
  author  = {Mehdi Bennis and M{\'e}rouane Debbah and H. Vincent Poor},
  title   = {Ultrareliable and Low-Latency Wireless Communication:
             Tail, Risk, and Scale},
  journal = {Proc. IEEE},
  year    = {2018},
  volume  = {106},
  number  = {10},
  pages   = {1834--1853},
  month   = oct,
  doi     = {10.1109/JPROC.2018.2867029}
}

@article{Lancho2023URLLC,
  author  = {Alejandro Lancho and Giuseppe Durisi and Luca Sanguinetti},
  title   = {Cell-Free Massive {MIMO} for {URLLC}:
             A Finite-Blocklength Analysis},
  journal = {IEEE Trans. Wireless Commun.},
  year    = {2023},
  volume  = {22},
  number  = {12},
  pages   = {8723--8735},
  month   = dec,
  doi     = {10.1109/TWC.2023.3265303}
}

@article{Chen2018Hardening,
  author  = {Zheng Chen and Emil Bj{\"o}rnson},
  title   = {Channel Hardening and Favorable Propagation in Cell-Free
             Massive {MIMO} With Stochastic Geometry},
  journal = {IEEE Trans. Commun.},
  year    = {2018},
  volume  = {66},
  number  = {11},
  pages   = {5205--5219},
  month   = nov,
  doi     = {10.1109/TCOMM.2018.2846272}
}

@article{Interdonato2019Training,
  author  = {Giovanni Interdonato and Hien Quoc Ngo and P{\aa}l Frenger and
             Erik G. Larsson},
  title   = {Downlink Training in Cell-Free Massive {MIMO}:
             A Blessing in Disguise},
  journal = {IEEE Trans. Wireless Commun.},
  year    = {2019},
  volume  = {18},
  number  = {11},
  pages   = {5153--5169},
  month   = nov,
  doi     = {10.1109/TWC.2019.2933831}
}

@article{VanTorre2011IndoorMIMO,
  author  = {Patrick Van Torre and Luigi Vallozzi and Carla Hertleer and
             Hendrik Rogier and Marc Moeneclaey and Jo Verhaevert},
  title   = {Indoor Off-Body Wireless {MIMO} Communication
             With Dual Polarized Textile Antennas},
  journal = {IEEE Trans. Antennas Propag.},
  year    = {2011},
  volume  = {59},
  number  = {2},
  pages   = {631--642},
  month   = feb,
  doi     = {10.1109/TAP.2010.2096389}
}

@article{Chong2024SINRDistribution,
  author  = {Baolin Chong and Fengqian Guo and Hancheng Lu and Langtian Qin},
  title   = {On the Distribution of {SINR} for Cell-Free Massive
             {MIMO} Systems},
  journal = {IEEE Trans. Commun.},
  year    = {2025},
  volume  = {73},
  number  = {2},
  pages   = {832--845},
  month   = feb,
  doi     = {10.1109/TCOMM.2024.3442694}
}

@article{Huang2025EEEURLLC,
  author  = {Yige Huang and Yanxiang Jiang and Fu-Chun Zheng and
             Pengcheng Zhu and Tony Q. S. Quek},
  title   = {Effective Energy Efficiency of Cell-Free {mMIMO} Systems
             for {URLLC} With Probabilistic Delay Bounds and
             Finite Blocklength Communications},
  journal = {IEEE Trans. Wireless Commun.},
  year    = {2025},
  volume  = {24},
  number  = {3},
  pages   = {2279--2296},
  month   = mar,
  doi     = {10.1109/TWC.2024.3519587}
}

@inproceedings{Chen2023LSFCMining,
  author    = {Yue Chen and Tao Peng and Yichen Guo and
               Chunmeng Fan and Wenbo Wang},
  title     = {Large-Scale Fading Coefficients Mining-Based Interference
               Identification and {SINR} Prediction for Cell-Free
               Massive {MIMO}},
  booktitle = {Proc. IEEE 98th Veh. Technol. Conf. (VTC2023-Fall)},
  address   = {Hong Kong},
  year      = {2023},
  month     = oct,
  pages     = {1--6},
  doi       = {10.1109/VTC2023-FALL60731.2023.10333684}
}

@article{Liu2026MLE,
  author  = {Longfei Liu and Yichen Guo and Yihang Cheng and
             Wenjie Li and Tao Peng and Wenbo Wang},
  title   = {A {MLE}-Based Joint Estimation Framework for
             Nakagami-$m$ Shape Parameter Estimation and
             Interference Identification in {HRLLC}},
  journal = {IEEE Wireless Commun. Lett.},
  year    = {2026},
  volume  = {15},
  pages   = {755--759},
  doi     = {10.1109/LWC.2025.3638778}
}

@article{Lin2021NakagamiCombiner,
  author  = {Jia-Chin Lin and H. Vincent Poor},
  title   = {Optimum Combiner for Spatially Correlated
             Nakagami-{$m$} Fading Channels},
  journal = {IEEE Trans. Wireless Commun.},
  year    = {2021},
  volume  = {20},
  number  = {2},
  pages   = {771--784},
  month   = feb,
  doi     = {10.1109/TWC.2020.3028284}
}

@inproceedings{Ngo2014Favorable,
  author    = {Hien Quoc Ngo and Erik G. Larsson and Thomas L. Marzetta},
  title     = {Aspects of Favorable Propagation in Massive {MIMO}},
  booktitle = {Proc. 22nd Eur. Signal Process. Conf. (EUSIPCO)},
  address   = {Lisbon, Portugal},
  year      = {2014},
  month     = sep,
  pages     = {76--80}
}

@article{Oehlert1992Delta,
  author  = {Gary W. Oehlert},
  title   = {A Note on the Delta Method},
  journal = {The American Statistician},
  year    = {1992},
  volume  = {46},
  number  = {1},
  pages   = {27--29},
  month   = feb,
  doi     = {10.1080/00031305.1992.10475842}
}
\vfill

\end{document}